\documentclass[aps,prd,onecolumn,groupedaddress,showpacs,nofootinbib,amssymb]{revtex4}
\usepackage[dvips]{graphicx}
\usepackage{amssymb}
\usepackage{amsmath}
\usepackage{graphicx,,color}
\usepackage{amsfonts}
\usepackage{bm}
\usepackage{cancel}
\usepackage{comment}

\newcommand\be{\begin{equation}}
\newcommand\ee{\end{equation}}

\allowdisplaybreaks[4]

\begin{document}

\tolerance=5000

\title{Constant-roll $k$-Inflation Dynamics}
\author{S.~D.~Odintsov,$^{1,2,3,4}$\,\thanks{odintsov@ieec.uab.es}
V.K.~Oikonomou,$^{5,6,7}$\,\thanks{v.k.oikonomou1979@gmail.com}}
\affiliation{$^{1)}$ ICREA, Passeig Luis Companys, 23, 08010 Barcelona, Spain\\
$^{2)}$ Institute of Space Sciences (IEEC-CSIC) C. Can Magrans
s/n,
08193 Barcelona, Spain\\
$^{3)}$ Tomsk State Pedagogical University, 634061 Tomsk, Russia\\
$^{4)}$ CAS Key Laboratory for Researches in Galaxies and
Cosmology, University of Science and Technology of China, Hefei, Anhui 230026, China\\
$^{5)}$ Department of Physics, Aristotle University of
Thessaloniki, Thessaloniki 54124,
Greece\\
$^{6)}$ Laboratory for Theoretical Cosmology, Tomsk State
University of Control Systems
and Radioelectronics, 634050 Tomsk, Russia (TUSUR)\\
$^{7)}$ Theoretical Astrophysics, IAAT, University of
T\"{u}bingen, Germany }

\tolerance=5000

\begin{abstract}
In this work we shall investigate the phenomenological
implications of the constant-roll condition on a $k$-Inflation
theory of gravity. The latter theories are particularly promising,
since these remained robust to the results of GW170817, since
these have a gravitational wave speed $c_T=1$ in natural units. We
shall mainly focus on the phenomenology of the $k$-Inflation
models, with the only assumption being the slow-roll condition
imposed on the first and  fourth slow-roll parameters, and the
constant-roll condition for the evolution of the scalar field. We
present in detail the final form of the gravitational equations of
motion that control the dynamics of the cosmological system, with
the constant-roll condition imposed, and by using a conveniently,
from the perspective of analytical manipulations, chosen
potential, we express the slow-roll indices and the resulting
observational indices of the theory as functions of the
$e$-foldings number. The results of our analysis indicate that the
constant-roll $k$-Inflation theory can be compatible with the
Planck 2018 data, for a wide range of the free parameters. Also we
examine in a quantitative way the effects of the constant-roll
condition on the parameter $f_{NL}^{equil}$ on which the
bispectrum is proportional, in the equilateral momentum
approximation, and we demonstrate that the effect of the
constant-roll condition is non-trivial. In effect,
non-Gaussianities in the theory may be enhanced, a phenomenon
which is known to be produced by constant-roll scalar theories of
gravity in general.
\end{abstract}


\maketitle

\section{Introduction}

One of the prominent problems in modern theoretical cosmology is
to describe accurately the primordial era of our Universe, and
particularly the inflationary era, during which our Universe
expanded in a nearly exponential rate and became homogenized. To
this problem, continuous efforts focusing on the Cosmic Microwave
Background structure and inhomogeneity, lead to astonishing
results that constrain quite strongly the inflationary spectrum to
great accuracy \cite{Akrami:2018odb}. Particularly, the power
spectrum of the scalar primordial curvature perturbations seems to
be nearly scale invariant and the tensor-to-scalar ratio is small
of the order $\mathcal{O}(10^{-2})$. Initially, inflationary
theories were materialized in terms of a scalar field, the
inflaton, which controlled the evolution of the Universe during
the rapid primordial accelerating era
\cite{Guth:1980zm,Starobinsky:1982ee,Linde:1983gd,Albrecht:1982wi},
however nowadays modified gravity also provides a consistent
theoretical framework that can harbor the inflationary era, for
reviews see
\cite{Nojiri:2017ncd,Nojiri:2010wj,Nojiri:2006ri,Capozziello:2011et,Capozziello:2010zz,delaCruzDombriz:2012xy,Olmo:2011uz}.
The Planck satellite observational data severely constrained the
inflationary era, and narrowed down to a great extent the viable
inflationary scenarios, leaving only a few scalar theories of
inflation intact with regards their viability. However, many
modified gravity models still remain quite viable and compatible
with the Planck predictions \cite{Akrami:2018odb}.

In 2017, the GW170817 event \cite{GBM:2017lvd} further constrained
the inflationary theories, due to the fact that the gravitational
waves and the electromagnetic signal emitted from the merging of
the two neutron stars came simultaneously. This indicated that the
gravitational wave speed was $c_T^2=1$ in natural units, thus this
phenomenal event eliminated many theoretical models of inflation,
like most of the Horndeski theories, and many of the string
corrected models involving couplings of the scalar field to the
Gauss-Bonnet invariant, see Ref. \cite{Ezquiaga:2017ekz} for
details on the currently allowed theories. One of the surviving
theories are the so-called $k$-Inflation theories
\cite{ArmendarizPicon:1999rj,Chiba:1999ka,ArmendarizPicon:2000dh,Matsumoto:2010uv,ArmendarizPicon:2000ah,Chiba:2002mw,Malquarti:2003nn,Malquarti:2003hn,Chimento:2003zf,Chimento:2003ta,Scherrer:2004au,Aguirregabiria:2004te,ArmendarizPicon:2005nz,Abramo:2005be,Rendall:2005fv,Bruneton:2006gf,dePutter:2007ny,Babichev:2007dw,Deffayet:2011gz,Kan:2018odq,Unnikrishnan:2012zu,Li:2012vta}
which have also the appealing feature of being able to describe
the dark energy era too.

In this paper we shall investigate the phenomenological features
of $k$-Inflation theories by assuming that the constant-roll
condition
\cite{Inoue:2001zt,Tsamis:2003px,Kinney:2005vj,Tzirakis:2007bf,
Namjoo:2012aa,Martin:2012pe,Motohashi:2014ppa,Cai:2016ngx,
Motohashi:2017aob,Hirano:2016gmv,Anguelova:2015dgt,Cook:2015hma,
Kumar:2015mfa,Odintsov:2017yud,Odintsov:2017qpp,Lin:2015fqa,Gao:2017uja,Nojiri:2017qvx,Oikonomou:2017bjx,Odintsov:2017hbk,Oikonomou:2017xik,Cicciarella:2017nls,Awad:2017ign,Anguelova:2017djf,Ito:2017bnn,Karam:2017rpw,Yi:2017mxs,Mohammadi:2018oku,Gao:2018tdb,Mohammadi:2018wfk,Morse:2018kda,Cruces:2018cvq,GalvezGhersi:2018haa,Boisseau:2018rgy,Gao:2019sbz,Lin:2019fcz}
holds true. The astonishing feature of the constant-roll
condition, is that it is associated with primordial
non-Gaussianities in the power spectrum of the CMB
\cite{Liddle:2000cg}. In ordinary $k$-Inflation theories, if the
slow-roll assumption is taken into account, the non-Gaussianities
are expected to be small \cite{DeFelice:2011zh}, however the
constant-roll condition may enhance the non-Gaussianities in the
bispectrum, in the equilateral momentum approximation. Our purpose
is to present the formalism of $k$-Inflation theories in the
constant-roll approximation in detail, and calculate the spectral
index of the scalar primordial curvature perturbations and the
tensor-to-scalar ratio. In addition, we shall investigate for
certain models the factor $f_{NL}^{equil}$ that appears in the
bispectrum, in the equilateral momentum approximation. The
$k$-Inflation theories are theories of the general form
$f(R,\phi,X)$, with
$X=\frac{1}{2}\partial^{\mu}\phi\partial_{\mu}\phi$, the
perturbations of which were studied in detail in
Refs.~\cite{Noh:2001ia,Hwang:2005hb,Hwang:2002fp,Kaiser:2013sna}.
Our aim is to present the general form of the ``slow-roll''
indices capturing the inflationary dynamics in the constant-roll
approximation, and investigate whether these theories can provide
a viable phenomenology, compatible with the latest Planck data
\cite{Akrami:2018odb}. In addition, we shall examine the
possibility whether the bispectrum can be enhanced by the
constant-roll condition. For the moment, the Gaussian structure of
the primordial scalar models of the perturbations is not
questioned, but future observations may distort the Gaussianity
assumption, so this paper aims to introduce a theoretical
description that may survive from the Planck, GW170817 and from
future observations that may indicate the presence of
non-Gaussianities in the power spectrum of the primordial scalar
perturbations modes. Apart from this fact which clearly motivates
the use of the constant-roll condition, we need to explain the
motivation for using a $k$-inflation theory to study its
phenomenological aspects. The motivation comes from the fact that
the $k$-inflation theory is the most general theory that can
describe inflation with a single scalar field in the context of
Einstein gravity \cite{Lyth:2009zz}. Apart from this, the GW170817
result significantly narrowed down the possible theories that may
consistently describe gravitational interactions, thus the
remaining viable theories must in some way studied in further
detail in order to reveal all the possible phenomenological
implications of these.

Our study will be focused on flat background geometries, and
specifically we shall assume that the background geometry is that
of a flat Friedman-Robertson-Walker (FRW), with line element,
\begin{equation}
\label{metricfrw}
ds^2 = - dt^2 + a(t)^2 \sum_{i=1,2,3} \left(dx^i\right)^2\, ,
\end{equation}
where $a(t)$ is the scale factor.

\section{Constant-roll $k$-Inflation Gravity: Equations of Motion and Inflationary Dynamics}

The $k$-Inflation model we shall consider, belongs to the general
class of $f(R,\phi,X)$ theories, with gravitational action,
\begin{equation}
\label{mainactionB} \mathcal{S}=\int d^4x\sqrt{-g}\left[
\frac{1}{2}f(R,\phi,X) \right]\, ,
\end{equation}
where in our case, the function $f(R,\phi,X)$ which we shall
consider is equal to,
\begin{equation}\label{frxfunction}
f(R,\phi,X)=\frac{R}{\kappa^2}-2\alpha X-2V(\phi)+\gamma X^2\, .
\end{equation}
In Eq. (\ref{frxfunction}),
$X=\frac{1}{2}\partial^{\mu}\phi\partial_{\mu}\phi$, $V(\phi)$ is
the scalar potential, $\alpha$ is a real number which will be
assumed to be equal to $\alpha=1$, with this value corresponding
to a canonical scalar field, however we keep the notation for
general $\alpha$ because in this way if one wants to study the
phantom scalar case, which corresponds to $\alpha=-1$, one just
have to replace $\alpha=-1$ in the resulting equations. Finally,
$\kappa^2=\frac{1}{M_p^2}$, where $M_p$ is the reduced Planck
mass, and $\gamma$ is a free parameter of mass units $[m]^{-4}$.
For the FRW background (\ref{metricfrw}), the gravitational
equations of motion become,
\begin{equation}\label{euqationsofmotion1}
3H^2=\frac{1}{F}(Xf_{,X}+\frac{RF-f}{2}-3H\dot{F})\, ,
\end{equation}
\begin{equation}\label{euqationsofmotion2}
-2\dot{H}-3H^2=\frac{1}{F}\left(
-\frac{RF-f}{2}+\ddot{F}+2H\dot{F} \right)\, ,
\end{equation}
\begin{equation}\label{euqationsofmotion3}
\frac{1}{a^3}\left( a^3\dot{\phi}f_{,X}\right)^{.}+f_{,\phi}=0\, ,
\end{equation}
where the ``dot'' indicates differentiation with respect to the
cosmic time, and $F=\frac{\partial f}{\partial R}$. Using the
functional form of the function $f(R,\phi,X)$ (\ref{frxfunction}),
the gravitational equations become,
\begin{equation}\label{euqationsofmotion11}
\frac{3H^2}{\kappa^2}=-\alpha X+\frac{3\gamma X^2}{4}+V(\phi)\, ,
\end{equation}
\begin{equation}\label{euqationsofmotion21}
-\frac{2\dot{H}+3H^2}{\kappa^2}=-\alpha X-V(\phi)+\frac{\gamma
X^2}{2}\, ,
\end{equation}
\begin{equation}\label{euqationsofmotion31}
3H\dot{\phi}(-2\alpha-\gamma\dot{\phi}^2)-2\gamma
\dot{\phi}^2\ddot{\phi}-(2\alpha+\gamma\dot{\phi}^2)\ddot{\phi}+2V'(\phi)=0\,
,
\end{equation}
and note that $F=\frac{1}{\kappa^2}$ in our case. The dynamical
evolution of the cosmological system is controlled by the
differential equations (\ref{euqationsofmotion11}),
(\ref{euqationsofmotion21}), (\ref{euqationsofmotion31}), given
the potential $V(\phi)$, however solving these analytically is a
formidable task, unless some simplification is implied. We shall
quantify the dynamics of inflation in terms of the ``slow-roll''
parameters \cite{Hwang:2005hb} (which are traditionally called
like this, without assuming for the moment that these are small
numbers),
\begin{align}\label{slowrollparameters}
& \epsilon_1=\frac{\dot{H}}{H^2}\, , \,\,\,
\epsilon_2=\frac{\ddot{\phi}}{H\dot{\phi}}\, ,\,\,\,
\epsilon_3=\frac{\dot{F}}{2HF}\, ,\,\,\,
\epsilon_4=\frac{\dot{E}}{2HE}\, ,
\end{align}
where $E$ in our case is,
\begin{equation}\label{epsilonE}
E=-\frac{F}{2X}\left(X f_{,X}+2X^2 f_{,XX}
\right)\, .
\end{equation}
Note here that we used a different definition of the slow-roll
$\epsilon_1$ in comparison to the standard one
\cite{Stewart:1993bc,Habib:2002yi}. Hereafter we impose the
slow-roll condition, only on the slow-roll parameters
$\epsilon_1=\frac{\dot{H}}{H^2}$ and $\epsilon_4$, so
$\epsilon_1,\epsilon_4 \ll 1$ during the inflationary era, and
also we assume that $\dot{\phi}^2\ll V(\phi)$. However, these
assumptions-constraints must be checked if they hold true. The
slow-roll condition only on $\epsilon_1$ is necessary in order to
ensure an exit from the inflationary era, when this slow-roll
parameter becomes of the order $\mathcal{O}(1)$. In addition, we
shall assume that the constant-roll condition holds true, which
is,
\begin{equation}\label{constantrollcondition}
\ddot{\phi}=\beta H \dot{\phi}\, ,
\end{equation}
where $\beta$ is some dimensionless real parameter of the order
$\mathcal{O}(1)$, so the constant-roll condition affects
practically the ``slow-roll'' parameter $\epsilon_2$ which is
required to be $\epsilon_2=\beta$.

For the FRW metric, assuming that the scalar field depends solely
on the cosmic time, $X$ becomes $X=-\frac{\dot{\phi}^2}{2}$, and
by substituting this in Eqs. (\ref{euqationsofmotion11}),
(\ref{euqationsofmotion21}), (\ref{euqationsofmotion31}), in
conjunction with the constant-roll condition
(\ref{constantrollcondition}), the gravitational equations of
motion become,
\begin{equation}\label{euqationsofmotion111}
\frac{3H^2}{\kappa^2}=\frac{\alpha}{2}\dot{\phi}^2+3\gamma
\frac{\dot{\phi}^4}{8}+V(\phi)\, ,
\end{equation}
\begin{equation}\label{euqationsofmotion211}
-\frac{2\dot{H}+3H^2}{\kappa^2}=\frac{\alpha}{2}\dot{\phi}^2+\gamma
\frac{\dot{\phi}^4}{8}-V(\phi)\, ,
\end{equation}
\begin{equation}\label{euqationsofmotion311}
-2\alpha
H\dot{\phi}(3+\beta)-3H\gamma\dot{\phi}^3(\beta+1)-2V'(\phi)=0\, .
\end{equation}
By taking into account the assumption $\frac{\dot{\phi}^2}{2}\ll
V(\phi)$, the first two gravitational equations become more
simple,
\begin{equation}\label{euqationsofmotion1111}
\frac{3H^2}{\kappa^2}\simeq V(\phi)\, ,
\end{equation}
\begin{equation}\label{euqationsofmotion2111}
\dot{H}\simeq -\kappa^2\left(\frac{\alpha}{4}\dot{\phi}^2+\gamma
\frac{\dot{\phi}^4}{16}\right)\, .
\end{equation}
From Eq. (\ref{euqationsofmotion311}) it is apparent that the
$k$-Inflation contribution to the inflationary evolution comes from
terms $\sim \dot{\phi}^3$.

Let us discuss certain features of Eq. (\ref{euqationsofmotion311}). It is apparently a third order equation with respect to $\dot{\phi}$, so it has three solutions, two of which are complex and thus not physically interesting, thus we shall use the one with real values of $\dot{\phi}$. An important comment is in order, notice that if we take the limit $\beta \to 0$ and $\gamma \to 0$ in Eq. (\ref{euqationsofmotion311}), we obtain,
\begin{equation}\label{slowrollsolution}
-3\alpha
H\dot{\phi}-V'(\phi)=0\, ,
\end{equation}
which is the slow-roll solution. If however one solves algebraically Eq. (\ref{euqationsofmotion311}), the limit $\beta \to 0$ and $\gamma \to 0$ cannot be taken in the solution, so the slow-roll solution cannot be obtained from the solutions of Eq. (\ref{euqationsofmotion311}) by taking the limit $\beta \to 0$ and $\gamma \to 0$. Mathematically this is easy to understand, since Eq. (\ref{euqationsofmotion311}) describes a curve with non-zero curvature as a function of $\dot{\phi}$, so the roots of Eq. (\ref{euqationsofmotion311}) are the points where this curve meets the $\dot{\phi}$ axis. The slow-roll solution describes a straight line, instead of a non-zero curvature curve in the plane. To see this more explicitly consider the curve,
\begin{equation}\label{curve}
\gamma +\beta  x^3+\alpha  x=0\, ,
\end{equation}
with $\alpha$, $\beta$ and $\gamma$ being real parameters. The above equation has three solutions, two of which are complex conjugate so we do not quote here, but the real solution is,
\begin{equation}\label{lastquote}
x=\frac{\sqrt[3]{\sqrt{3} \sqrt{\beta ^3 \left(4 \alpha ^3+27 \beta  \gamma ^2\right)}-9 \beta ^2 \gamma } \left(\frac{\sqrt[3]{2}}{\beta }-\frac{2 \sqrt[3]{3} \alpha }{\left(\sqrt{3} \sqrt{\beta ^3 \left(4 \alpha ^3+27 \beta  \gamma ^2\right)}-9 \beta ^2 \gamma \right)^{2/3}}\right)}{6^{2/3}}\, .
\end{equation}
As it can be seen, we cannot take the limit $\beta \to 0$, infinities occur, but this limit can be taken for the algebraic equation (\ref{curve}), yielding $\gamma +\alpha  x=0$. Let us return to the problem at hand, so by solving the algebraic equation
(\ref{euqationsofmotion311}) with respect to $\dot{\phi}$, we
obtain the following real solution,
\begin{equation}\label{dotphi1}
\dot{\phi}\simeq \frac{6 \alpha  (\beta +1) (\beta +3) \gamma  \kappa ^2 V(\phi )-\left(81 \Delta (\phi )+9 \sqrt{S(\phi )}\right)^{2/3}}{3\ 3^{5/6} (\beta +1) \gamma  \kappa  \sqrt{V(\phi )} \sqrt[3]{81 \Delta (\phi )+9 \sqrt{S(\phi )}}}
\, ,
\end{equation}
and by substituting this to Eq. (\ref{constantrollcondition}) in
conjunction with Eq. (\ref{euqationsofmotion1111}) we get,
\begin{equation}\label{ddotphi1}
\ddot{\phi}=\beta  \kappa \sqrt{\frac{V(\phi)}{3}}  \left(\frac{6 \alpha  (\beta +1) (\beta +3) \gamma  \kappa ^2 V(\phi )-\left(81 \Delta (\phi )+9 \sqrt{S(\phi )}\right)^{2/3}}{3\ 3^{5/6} (\beta +1) \gamma  \kappa  \sqrt{V(\phi )} \sqrt[3]{81 \Delta (\phi )+9 \sqrt{S(\phi )}}} \right )\, ,
\end{equation}
where $S(\phi)$ and $\Delta (\phi)$ in both Eqs. (\ref{dotphi1})
and (\ref{ddotphi1}) are defined as follows,
\begin{equation}\label{defintions1}
S(\phi)=(\beta +1)^3 \gamma ^3 \kappa ^4 V(\phi )^2 \left(81
(\beta +1) \gamma  V'(\phi )^2+\frac{8}{3} \alpha ^3 (\beta +3)^3
\kappa ^2 V(\phi )\right)\, ,
\end{equation}
\begin{equation}\label{defintions2}
\Delta(\phi)=(\beta +1)^2 \gamma ^2 \kappa ^2 V(\phi ) V'(\phi )\,
.
\end{equation}
We also have the wave speed which characterizes the propagation of
the primordial perturbations, for the $k$-Inflation theory, which
will be needed for the calculation of the tensor-to-scalar ratio,
\begin{equation}\label{cA1}
c_A^2=\frac{f_{,X}}{f_{,X}+2Xf_{,XX}}\, ,
\end{equation}
while the gravitational wave speed is $c_T=1$, which is why the
$k$-Inflation theories are still compatible with the GW170817 event.

Let us proceed to give the expressions of the ``slow-roll''
indices for the theory at hand, having in mind that these must be
evaluated at the horizon crossing time instance, where the value
of the scalar field is $\phi=\phi_k$. After some simple
calculations and by combining Eqs. (\ref{slowrollparameters}),
(\ref{epsilonE}), (\ref{euqationsofmotion1111}) and
(\ref{euqationsofmotion2111}), these are equal to,
\begin{align}\label{slowrollparametersanalytic}
& \epsilon_1(\phi)=-\frac{3\left(
\frac{\alpha}{4}\dot{\phi}^2+\gamma
\frac{\dot{\phi}^4}{16}\right)\kappa^2}{V(\phi)}\, , \\ \notag &
\epsilon_2(\phi)=\beta\, , \\ \notag & \epsilon_3(\phi)=0\, ,\\
\notag & \epsilon_4(\phi)=\frac{3\sqrt{3}\gamma
\dot{\phi}\ddot{\phi}}{\kappa\sqrt{V(\phi)}\left(2\alpha+3\gamma\dot{\phi}^2
\right)}\, ,
\end{align}
where $\dot{\phi}$ and $\ddot{\phi}$ are functions of the scalar
field $\phi$, as in Eqs. (\ref{dotphi1}) and (\ref{ddotphi1}).
Accordingly, the spectral index of the primordial curvature
perturbations is equal to \cite{Hwang:2005hb},
\begin{equation}\label{spectralindex1}
n_s=1+2\frac{2\epsilon_1-\epsilon_2+\epsilon_3-\epsilon_4}{1+\epsilon_1}\,
,
\end{equation}
which holds true even in the case that the $\epsilon_i$ do not
take small values. In addition, the analytic form of the
tensor-to-scalar ratio $r$ is for the theory at hand
\cite{Hwang:2005hb},
\begin{equation}\label{tensortoscalarratio1}
r=4\left(\frac{\Gamma (3/2)}{\Gamma
(3/2+\epsilon_2)2^{\epsilon_2}}c_A^{3/2+\epsilon_2}
\frac{\sqrt{3}\dot{\phi}\sqrt{(2\alpha+3\gamma
\dot{\phi}^2)}}{\sqrt{2V(\phi)}} \right)^2
\end{equation}
which must be evaluated at the horizon crossing. For the above
calculation we assumed that the slow-roll indices $\epsilon_1$ and
$\epsilon_4$ take small values during the inflationary era, a
condition which we must check if it holds true at the end of this
section. Notice that for the slow-roll case, the tensor-to-scalar
ratio is $r=4 |\epsilon_1| c_A$, so the constant-roll and
$k$-Inflation effects are materialized in the parameters $\beta$
and $\gamma$ in Eq. (\ref{tensortoscalarratio1}). Having the
slow-roll indices as functions of the scalar field, now we can
proceed in expressing those as functions of the $e$-foldings
number $N$, which is defined as follows,
\begin{equation}\label{efoldings1cosmictime}
N=\int_{t_i}^{t_f}H(t)dt\, ,
\end{equation}
where $t_i$ is the time instance that inflation starts, which we
shall assume it to be equal to the time instance of horizon
crossing, at which $k=aH$, while $t_f$ is the time instance that
inflation ends. Of course inflation is known to start quite
earlier than the horizon crossing time instance, but we choose
$t_i$ to be the horizon crossing time instance, because the
condition $k=aH$ is vital for the calculation of the power
spectrum, both scalar and tensor. Also we need to note that the
initial condition of the constant-roll $k$-inflation theory may be
a Bunch-Davies vacuum state \cite{Oikonomou:2017isf} or a quasi-de
Sitter like expansion which leads to a Bunch-Davies vacuum state
\cite{Keskin:2018gev}. Expressed in terms of the scalar field, the
$e$-foldings number is equal to,
\begin{equation}\label{efoldings2}
N=\int_{\phi_k}^{\phi_f}\frac{H}{\dot{\phi}}d\phi\, ,
\end{equation}
or by using Eq. (\ref{euqationsofmotion1111}), this can be written
as,
\begin{equation}\label{efoldings3}
N=\int_{\phi_k}^{\phi_f}
\frac{\sqrt{\frac{V(\phi)}{3}}\kappa}{\dot{\phi}}\, ,
\end{equation}
which in view of Eq. (\ref{dotphi1}) yields the $e$-foldings
number as a function of the scalar field $\phi$. The value of the
scalar field $\phi_f$ at the end of the inflationary era can be
evaluated easily, since at that era, $\epsilon_1(\phi_f)\sim
\mathcal{O}(1)$, so $\phi_f$ is obtained. Thus by substituting the
result in Eq. (\ref{efoldings3}) and performing the
$\phi$-integration, one can have $\phi_k$ as a function of the
$e$-foldings number. In effect, the slow-roll indices
(\ref{slowrollparametersanalytic}) and the corresponding
observational indices (\ref{spectralindex1}) and
(\ref{tensortoscalarratio1}) can be expressed as functions of the
$e$-foldings number, and a direct confrontation with the
observational data can be done. In the next section we shall
examine some specific examples of potentials, in order to see
whether a viable phenomenology can be obtained from the
constant-roll $k$-Inflation theory.

\subsection{Inflationary Phenomenology of Constant-roll $k$-Inflation
Power-law Scalar Potentials}

In the previous subsection we developed the formalism for the
inflationary dynamics of constant-roll $k$-Inflation theory, and in
this section we shall employ it in order to investigate the
phenomenology of power-law potentials. However, the most serious
setback is the lack of analyticity, so let us choose a relatively
simple  power-law potential,
\begin{equation}\label{potential}
V(\phi)=V_0\phi^n\, ,
\end{equation}
where $V_0$ is some positive parameter with mass dimensions
$[m]^{4-n}$. For this potential, it is easy to obtain the Hubble
rate and $\dot{\phi}$ as functions of $\phi$ in closed form,
however, in order to find analytically the value of the scalar
field at the end of inflation $\phi_f$ and eventually to express
the value of the scalar field at horizon crossing $\phi_k$ as a
function of the $e$-foldings number $N$, we shall assume that the
term $81 (\beta +1) \gamma  V'(\phi )^2$ is much larger than
$\frac{8}{3} \alpha ^3 (\beta +3)^3 \kappa ^2 V(\phi )$, that is,
\begin{equation}\label{assumption}
81 (\beta +1) \gamma  V'(\phi )^2\gg \frac{8}{3} \alpha ^3 (\beta
+3)^3 \kappa ^2 V(\phi )\, ,
\end{equation}
with both the terms appearing in the function $S(\phi)$ in Eq.
(\ref{dotphi1}), a condition that can be achieved if $\gamma \gg
\kappa $. Let us note here that the condition (\ref{assumption})
is not related to the condition $\frac{\dot{\phi}^2}{2}\ll
V(\phi)$. However, we impose the condition (\ref{assumption}) in
order to make the analytic treatment easier, and we need to check
after we present the phenomenology of the model, whether this
holds true for the values of the free parameters that guarantee
the phenomenological viability of the model.

Hereafter we choose reduced Planck units, in which $\kappa^2=1$,
so the only assumption we shall make regarding the free parameters
is that $\gamma\gg 1$ in Planck units. Having these assumptions in
mind, $\dot{\phi}$ can be evaluated in a simplified way and it is
equal to,
\begin{equation}\label{dotphiapprox}
\dot{\phi}\simeq -\frac{\sqrt[3]{2} \sqrt[3]{n} \sqrt[6]{V_0} \phi ^{\frac{n-2}{6}}}{\sqrt[6]{3} \sqrt[3]{\beta +1} \sqrt[3]{\gamma } \sqrt[3]{\kappa }}\, .
\end{equation}
Also, the value of the scalar field at the end of inflation
$\phi_f$ can be evaluated by taking $\epsilon_1(\phi)$ appearing
in Eq. (\ref{slowrollparametersanalytic}) to be equal to
$\epsilon_1(\phi_f)=1$, so by combining Eqs. (\ref{dotphiapprox})
and (\ref{potential}), we obtain the approximate solution,
\begin{equation}\label{approxphifinal}
\phi_f\simeq 2^{-\frac{2}{n+1}} 3^{-\frac{2}{3 \left(-\frac{2
n}{3}-\frac{2}{3}\right)}} \left(\frac{(\beta +1)^{2/3}
\gamma^{2/3} \kappa ^{2/3} V_0^{2/3}}{\alpha
n^{2/3}}\right)^{-\frac{3}{2(n+1)}}\, ,
\end{equation}
and notice that in the end we must take $\kappa =1$ in reduced
Planck units. With $\phi_f$, $\dot{\phi}$ and the potential given,
we can combine these and substitute their value in Eq.
(\ref{efoldings3}) in order to express the value of the scalar
field at horizon crossing as a function of the $e$-foldings
number, and by doing so we obtain,
\begin{align}\label{phikathorizoncrossing}
& \phi_k\simeq \left(\frac{2}{9}\right)^{\frac{1}{n+4}}
\left(\frac{\sqrt[3]{n} (n+4) \left(\frac{3^{2/3} \sqrt[3]{\beta
+1} \sqrt[3]{\gamma } \kappa ^{4/3} \sqrt[3]{V_0}
\left(\frac{\left(\frac{3}{4}\right)^{\frac{1}{n+1}} (\beta
+1)^{-\frac{1}{n+1}} \gamma ^{-\frac{1}{n+1}} \kappa
^{-\frac{1}{n+1}} V_0^{-\frac{1}{n+1}}}{\alpha^{-\frac{3}{2(n+1)}}
n^{-\frac{1}{n+1}}}\right)^{\frac{n+4}{3}}}{\sqrt[3]{2}
\sqrt[3]{n} (n+4)}+N\right)}{\sqrt[3]{\beta +1} \sqrt[3]{\gamma }
\kappa ^{4/3} \sqrt[3]{V_0}}\right)^{\frac{3}{n+4}}\, .
\end{align}
Having the above at hand, we can express the slow-roll indices
(\ref{slowrollparametersanalytic}) and the corresponding
observational indices (\ref{spectralindex1}) and
(\ref{tensortoscalarratio1}), as functions of the $e$-foldings
number, and we can consequently confront the theory with the
observational data. The 2018 Planck data \cite{Akrami:2018odb}
constrain the spectral index and the tensor-to-scalar ratio as
follows,
\begin{equation}\label{planck2018}
n_s= 0.9649 \pm 0.0042,\,\,\,r<0.064\, ,
\end{equation}
so let us investigate the phenomenology of the constant-roll
$k$-Inflation model. The resulting expressions for the observational
indices are quite large to quote these here, but we shall quote
the results of our analysis.

\begin{figure}[h!]
\centering
\includegraphics[width=18pc]{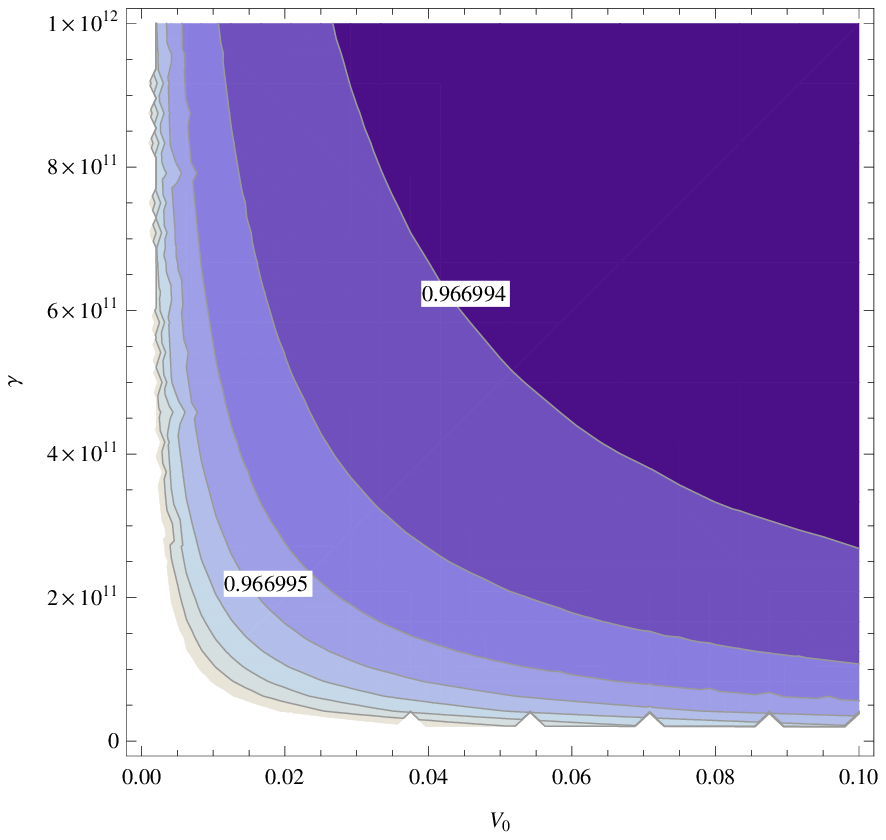}
\includegraphics[width=18pc]{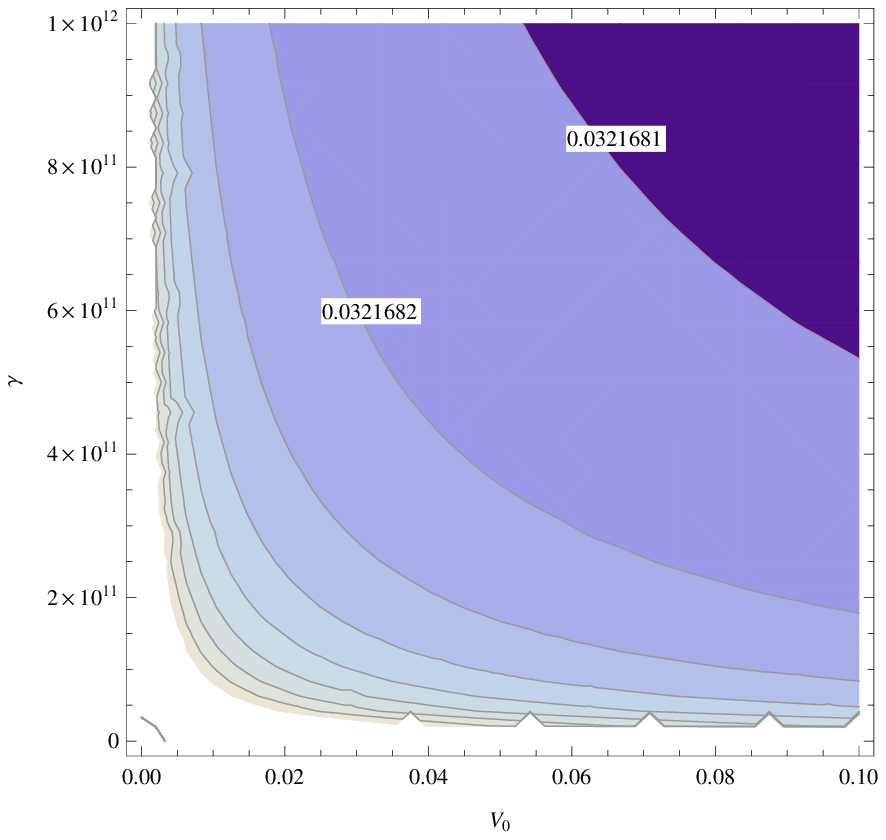}
\caption{The contour plot of $n_s$ (left plot) and $r$ (right
plot), for $(N,\beta ,\alpha ,\kappa ,n)=(60,0.007 ,1, 1
,n=1.00006)$, for $V_0$ chosen in the range $V_0=[0,0.1]$ and
$\gamma=[0, 10^{12}]$. In both plots, the lighter colors indicate
larger values of the plotted quantities, and darker colors
indicate smaller values of the respective plotted quantity. In
both plots we indicated some characteristic values of the
respective plotted quantity.} \label{plot1}
\end{figure}
\begin{figure}[h!]
\centering
\includegraphics[width=18pc]{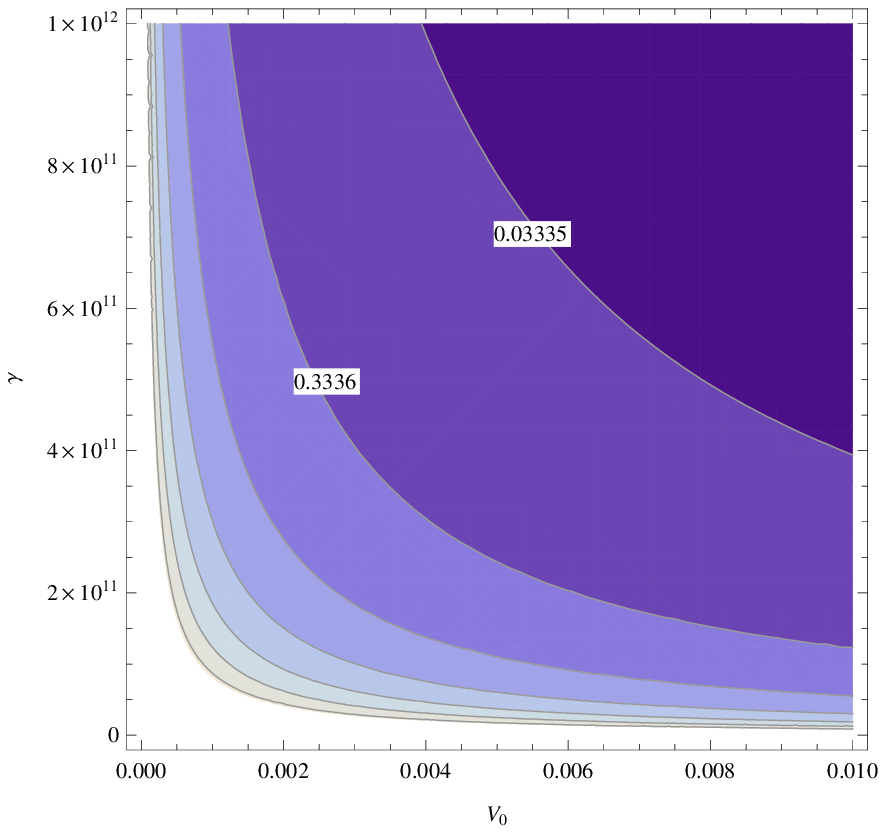}
\includegraphics[width=18pc]{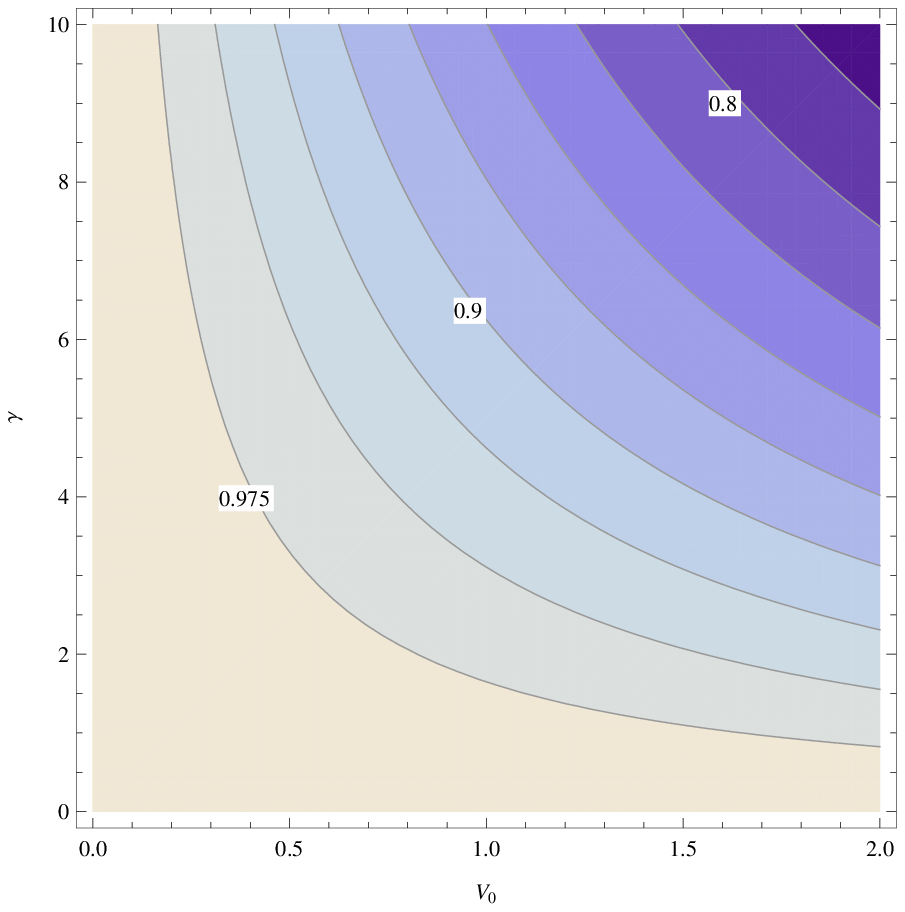}
\caption{The contour plot of $c_A^2$ for $(N,\beta ,\alpha ,\kappa
,n)=(60,0.007 ,1, 1 ,1.00006)$, for various values of $V_0$ and
$\gamma$. As it can be seen, the values of $c_A^2$ never exceed
unity and for the values of $V_0$ and $\gamma$ for which the
compatibility of the observational indices of inflation with the
observational data is achieved, we have $c_A^2\simeq 1$.}
\label{plot2}
\end{figure}

The compatibility of the resulting theory with the observational
data can be achieved only for small values of the parameter $\beta
$ and actually, the parameter $\beta$ crucially affects the
viability of the model. Particularly, it seems that $\beta$
strongly alters the phenomenology of the model and the rest of the
parameters have marginal effects, apart from the parameter $n$
which strongly affects the tensor-to-scalar ratio. For example for
$V_0=0.1$, $\gamma =10^{11.439998}$ in reduced Planck units, and
for $\beta=0.007$, $\alpha =1$ and $n=1.00006$, we get the
following values for the observational indices,
\begin{equation}\label{observationaldatanewresults}
n_s=0.966994,\,\,\,r=0.0321681\, ,
\end{equation}
which are compatible with the latest Planck data
(\ref{planck2018}). Recall that the 2018 Planck data
(\ref{planck2018}) indicate that the spectral index should be in
the range $n_s=[0.9607,0.9691]$, so clearly the results
(\ref{observationaldatanewresults}) are within the allowed range.
In addition, by choosing for example, $V_0=10^{10}$, $\gamma
=10^{11.439998}$ in reduced Planck units, and for $\beta=0.0059$,
$\alpha =1$ and $n=20$, we get the following values for the
observational indices,
\begin{equation}\label{observationaldatanewresults121}
n_s=0.955457,\,\,\,r=0.134348\, ,
\end{equation}
so it is obvious that $V_0$, $n$ and $\gamma$ do not affect the
spectral index, however, the tensor-to-scalar ratio is strongly
affected by the parameter $n$. Indeed, if we choose $V_0=10^{10}$,
$\gamma =10^{11.439998}$ in reduced Planck units, and for
$\beta=0.0059$, $\alpha =1$, as in the previous case, and choose
additionally $n=1$, the tensor-to-scalar ratio drops drastically
to $r=0.032$. So it seems that only the parameters $\beta$ and $n$
have a strong effect on the observational indices. In order to
better illustrate this issue, in Fig. \ref{plot1} we present the
contour plots of the spectral index $n_s$ (left plot) and of the
tensor-to-scalar ratio (right plot) as functions of $V_0$ and
$\gamma$, for $(N,\beta ,\alpha ,\kappa,n)=(60,0.007 ,1, 1
,1.00006)$, and for $V_0$ chosen in the range $V_0=[0,0.1]$ and
$\gamma=[0, 10^{12}]$. In both the plots it is apparent that $V_0$
and $\gamma$ do not crucially affect the observational indices. In
both the plots, the lighter colors indicate larger values of the
plotted quantities, and darker colors indicate smaller values of
the respective plotted quantity. In both plots we indicated some
characteristic values of the respective plotted quantity.
Before closing it is worth discussing another important issue,
related with the values of the parameter $c_A^2$ which can be
related with Jeans instabilities if $c_A^2<0$ or it may have
superluminal values \cite{Babichev:2007dw}. In our case, the
results are free from superluminal velocities and Jeans
instabilities, since $c_A^2$ defined in Eq. (\ref{cA1}) takes
values $c_A^2<1$ for a large range of the values of the free
parameters chosen in such a way so that the observational indices
are compatible with the observational data. For example, if we
choose, $V_0=0.1$, $\gamma =10^{11.439998}$ in reduced Planck
units, and for $\alpha =1$ and $n=1.00006$, we get,
\begin{equation}\label{caaa}
c_A^2\simeq 0.577417\, .
\end{equation}
In order to better understand the behavior of the velocity
$c_A^2$, in Fig. \ref{plot2} we present the contour plots of
$c_A^2$ as a function of $V_0$ and $\gamma$.
As it can be seen from both plots of Fig. \ref{plot2}, the sound
wave speed $c_A^2$ hardly changes as $V_0$ and $\gamma$ take
different values. Furthermore, it can be shown that $n$ and
$\beta$ also do not drastically affect the sound wave speed
$c_A^2$, and in all cases, the wave speed takes positive values
with $c_A^2<1$. In this case too, darker colors indicate larger
values of the wave speed $c_A^2$, and in both plots we indicated
some characteristic values.

It is important to validate that the assumption
$\epsilon_1,\epsilon_4\ll 1$ we made in order to calculate the
tensor-to-scalar ratio, holds true, for the allowed values of the
free parameters of the model. Indeed, it can be checked that both
$\epsilon_1$, and $\epsilon_4$ take small values for the allowed
values of the free parameters, for example by choosing $V_0=0.1$,
$\gamma =10^{11.439998}$ in reduced Planck units, and for
$\beta=0.007$, $\alpha =1$ and $n=1.00006$, we have,
\begin{equation}\label{epsilon1epsilon4}
|\epsilon_1|\simeq 0.00124166\, , \,\,\,|\epsilon_4 |\sim
0.00699977\, ,
\end{equation}
which are indeed quite smaller than unity. For these indices, the
same rules apply, so these are crucially affected by the parameter
$\beta$.

Finally, it is also vital for the self-consistency of the model
and the results presented above, to check whether the condition
(\ref{assumption}) holds true for the values of the free
parameters that guarantee the phenomenological viability of the
model. By choosing for example, $V_0=0.1$, $\gamma
=10^{11.439998}$ in reduced Planck units, and for $\beta=0.007$,
$\alpha =1$ and $n=1.00006$, we have,
\begin{equation}\label{assumptioncheck1}
81 (\beta +1) \gamma  V'(\phi )^2=2.24635\times 10^{11}\, ,
\end{equation}
and
\begin{equation}\label{assumptioncheck2}
\frac{8}{3} \alpha ^3 (\beta +3)^3 \kappa ^2 V(\phi )=1.34081\, ,
\end{equation}
in reduced Planck units, which clearly indicates that the imposed
condition (\ref{assumption}) holds true.
\subsection{The non-Gaussianities Issue}

The primordial power spectrum of the curvature perturbations seems
up to date to be Gaussian, to a high accuracy. However, the future
observations may reveal non-Gaussianities in the spectrum, and
these are quantified in the bispectrum and trispectrum, which in
turn are quantified in the correlation functions $\langle
g_{k_1},g_{k_2}g_{k_3} \rangle\sim B_g(k_1+k_2+k_3)$ and $\langle
g_{k_1},g_{k_2}g_{k_3}g_{k_4} \rangle\sim T_g(k_1+k_2+k_3+k_4)$
\cite{Lyth:2009zz}. The momenta have to add up to zero, while the
function $B_g(k_1+k_2+k_3)$ is the bispectrum and
$T_g(k_1+k_2+k_3+k_4)$ is called the trispectrum. The
constant-roll condition is known to produce non-Gaussianities in
canonical single scalar field theory, so one of the purposes of
this paper is to investigate whether this is possible in the
context of constant-roll $k$-Inflation theory. The calculation for
the bispectrum $B_g(k_1+k_2+k_3)\sim f_{NL}^{equil}$ in classes of
$f(R,\phi,X)$ theories were performed in several works, but in
Ref. \cite{DeFelice:2011zh} the bispectrum was calculated without
assuming the slow-roll conditions on the slow-roll indices, so the
result fits in an optimal way our work, because in our case the
second slow-roll index does not take small values. We shall change
the notation of Ref. \cite{DeFelice:2011zh} in order to comply
with the previous sections of our work, so in the equilateral
momentum approximation, the parameter $f_{NL}^{equil}$ in terms of
the slow-roll indices reads,
\begin{equation}\label{fNL1}
f_{NL}^{equil}=\frac{40}{9 \kappa ^2
Q}\left.\frac{\mathcal{C}_1}{12}+\frac{17 \mathcal{C}_2}{96
c_A^2}+\frac{1}{72} \mathcal{C}_3 (H \kappa )-\frac{1}{24}
\mathcal{C}_4 \left(\kappa ^2 Q\right)-\frac{1}{24} \mathcal{C}_5
\left(\kappa ^4 Q^2\right)\right.\, ,
\end{equation}
where the parameters $Q$ and $\mathcal{C}_i$ are defined as
follows,
\begin{align}\label{parametersQCi}
& Q=\frac{w_1 \left(4 w_1 w_3+9 w_2^2\right)}{3 w_2^2}\, , \\
\notag & \mathcal{C}_1=-\frac{\epsilon_1 \left(3 c_A^2-2
\epsilon_1-\epsilon_2-3\right)}{c_A^4}\, ,
\\ \notag &
\mathcal{C}_2=-\frac{\epsilon_1 \left(-c_A^2-2
s+\epsilon_2+1\right)}{c_A^2}\, , \\ \notag &
\mathcal{C}_3=-\frac{\kappa
\left(\left(1-c_A^2\right) \Sigma +2 \lambda \right)}{H^3}\, , \\
\notag & \mathcal{C}_4=\frac{2 \epsilon_1}{c_A^2}\, , \\
\notag & \mathcal{C}_5=-\frac{\epsilon_1}{4} \, ,
\end{align}
and $c_A^2$ is defined in Eq. (\ref{cA1}), while the parameters
$w_i$ appearing in Eq. (\ref{parametersQCi}) are defined as
follows,
\begin{equation}\label{parameterswi}
w_1=w_2=\frac{1}{\kappa ^4}\, , \,\,\, w_2=\frac{2 H}{\kappa ^2}\,
, \,\,\,w_3=3 \Sigma -\frac{9 H^2}{\kappa ^2}\, .
\end{equation}
Finally $\Sigma$ appearing in Eq. (\ref{parameterswi}) is defined
as follows,
\begin{equation}\label{sigma}
\Sigma=\frac{1}{2}Xf_{,X}+X^2f_{,XX}=-2\alpha X+4\gamma X^2\, .
\end{equation}
By inserting the $C_i$'s and the $w_i$'s in Eq. (\ref{fNL1}), the
parameter $f_{NL}^{equil}$ reads,
\begin{equation}\label{fNL2}
f_{NL}^{equil}=-\frac{35 H^2 \epsilon_1}{108 c_A^2 \kappa ^2
\Sigma }-\frac{85 s}{54 c_A^2}-\frac{10 \epsilon_1}{9
c_A^2}+\frac{5 \epsilon_2}{12 c_A^2}+\frac{5
c_A^2}{81}-\frac{35}{108 c_A^2}+\frac{5 \kappa ^2 \Sigma
\epsilon_1}{108 H^2}-\frac{10 \lambda }{81 \Sigma }-\frac{5}{81}\,
.
\end{equation}
Finally, the parameters $s$ and $\lambda$ were introduced for
notational convenience, and these are defined as follows,
\begin{align}\label{parameterslambdaands}
s=\frac{\dot{c}_A}{Hc_A}\, , \,\,\,
\lambda=\frac{1}{2}X^2f_{,XX}\, .
\end{align}
In order to have a concrete idea on the new effects that the
constant-roll condition brings along in the $k$-Inflation theory,
let us use the numerical values of the free parameters for which
we achieved the compatibility with the observational data, and let
us compare the results with the slow-roll case. Obviously, the
non-Gaussianity will be enhanced by the presence of the term
$\frac{5 \epsilon_2}{12 c_A^2}$, since $\epsilon_2=\beta$. So for
$(N,V_0,\gamma ,\alpha , \beta ,\kappa ,n)=(60,0.1,10^{11.439998}
,1 ,0.007, 1 ,1.00006)$, we have, $\frac{5 \epsilon_2}{12
c_A^2}\sim 0.0087256$. This can be compared to the term $ \frac{10
\epsilon_1}{9 c_A^2}$ which for $(N,V_0,\gamma ,\alpha , \beta
,\kappa ,n)=(60,0.1,10^{11.439998} ,1 ,0.007, 1 ,1.00006)$ is
approximately $\frac{10 \epsilon_1}{9 c_A^2}\sim 0.00155172$,
which is nearly one order smaller in comparison to the term $\sim
\epsilon_2$. The result is still small, but more enhanced in
comparison to the slow-roll case, and also it is within the
acceptable limits of the 2018 Planck constraints on primordial
non-Gaussianities, which indicate that $f_{NL}^{equil}=-26\pm 47$
\cite{Akrami:2019izv}.

Thus the constant-roll condition may enhance the non-Gaussianities
in the power spectrum of primordial curvature perturbations of the
$k$-Inflation theory with power-law potential. In principle, more
phenomenologically interesting potentials can be used, but the
lack of analyticity restricted us to use the simplest choices.

\section{Conclusions}

In this paper we investigated the phenomenological implications of
the constant-roll condition in a $k$-Inflation theory in the
presence of scalar potential. We presented in detail the structure
of the gravitational equations of motion, in view of the
constant-roll condition, and by assuming that only the first
slow-roll index takes small values, thus quantifying the slow-roll
condition only on this index, we formed the set of differential
equations that governs the constant-roll $k$-Inflation theory. By
choosing a convenient potential, that may allow analytical
manipulation of the equations to some extend, we expressed the
slow-roll indices and the corresponding observational indices as
functions of the $e$-foldings number. As we demonstrated, the
observational indices of the resulting quadratic potential theory
can be compatible with the observational data coming from the
latest Planck 2018 collaboration. Also we examined in some detail
the quantitative effects of the constant-roll condition on the
parameter $f_{NL}^{equil}$ appearing in the bispectrum, in the
equilateral momentum approximation, and we demonstrated that
non-trivial effects occur, thus non-Gaussianities are enhanced. In
general, $f(R,X,\phi)$ theories of $k$-Inflation type, and
generalizations, are particularly useful for providing theoretical
descriptions both compatible with the Planck data on inflation and
more importantly with the striking GW170817 event. Work is in
progress towards a unified modified gravity $k$-Inflation theory
with general scalar potentials.

Finally we need to stress one more the fact that $k$-inflation
theories lead to a $c_T^2=1$ gravitational wave speed in natural
units \cite{Hwang:2005hb}, regardless of the values of the free
parameters of the theory. This was partially the motivation for
studying this specific class of theory with regard to its
phenomenological implications.

\section*{Acknowledgments}

This work is supported by MINECO (Spain), FIS2016-76363-P, and by
project 2017 SGR247 (AGAUR, Catalonia) (S.D.O). SDO is grateful to
Yi-Fu Cai for kind hospitality at USTC. This work is supported by
the DAAD program ``Hochschulpartnerschaften mit Griechenland
2016'' (Projekt 57340132) (V.K.O). V.K.O is indebted to Prof. K.
Kokkotas for his hospitality in the IAAT, University of
T\"{u}bingen.


\begin{thebibliography}{99}



\bibitem{Akrami:2018odb}
Y.~Akrami {\it et al.} [Planck Collaboration],
arXiv:1807.06211 [astro-ph.CO].



\bibitem{Guth:1980zm}
A.~H.~Guth,
Phys.\ Rev.\ D {\bf 23} (1981) 347. doi:10.1103/PhysRevD.23.347

\bibitem{Starobinsky:1982ee}
A.~A.~Starobinsky,
Phys.\ Lett.\ {\bf 91B} (1980) 99.
doi:10.1016/0370-2693(80)90670-X

\bibitem{Linde:1983gd}
A.~D.~Linde,
Phys.\ Lett.\ {\bf 129B} (1983) 177.
doi:10.1016/0370-2693(83)90837-7


\bibitem{Albrecht:1982wi}
A.~Albrecht and P.~J.~Steinhardt,
Phys.\ Rev.\ Lett.\  {\bf 48} (1982) 1220 [Adv.\ Ser.\ Astrophys.\
Cosmol.\  {\bf 3} (1987) 158]. doi:10.1103/PhysRevLett.48.1220



\bibitem{Nojiri:2017ncd}
S.~Nojiri, S.~D.~Odintsov and V.~K.~Oikonomou,
Phys.\ Rept.\ {\bf 692} (2017) 1 doi:10.1016/j.physrep.2017.06.001
[arXiv:1705.11098 [gr-qc]].

\bibitem{Nojiri:2010wj}
S.~Nojiri and S.~D.~Odintsov,
Phys.\ Rept.\ {\bf 505} (2011) 59
doi:10.1016/j.physrep.2011.04.001 [arXiv:1011.0544 [gr-qc]].

\bibitem{Nojiri:2006ri}
S.~Nojiri and S.~D.~Odintsov,
eConf C {\bf 0602061} (2006) 06
 [Int.\ J.\ Geom.\ Meth.\ Mod.\ Phys.\ {\bf 4} (2007) 115]
doi:10.1142/S0219887807001928 [hep-th/0601213].

\bibitem{Capozziello:2011et}
S.~Capozziello and M.~De Laurentis,
Phys.\ Rept.\ {\bf 509} (2011) 167
doi:10.1016/j.physrep.2011.09.003 [arXiv:1108.6266 [gr-qc]].

\bibitem{Capozziello:2010zz}
V.~Faraoni and S.~Capozziello,
Fundam.\ Theor.\ Phys.\ {\bf 170} (2010).
doi:10.1007/978-94-007-0165-6

\bibitem{delaCruzDombriz:2012xy}
A.~de la Cruz-Dombriz and D.~Saez-Gomez,
Entropy {\bf 14} (2012) 1717 doi:10.3390/e14091717
[arXiv:1207.2663 [gr-qc]].

\bibitem{Olmo:2011uz}
G.~J.~Olmo,
Int.\ J.\ Mod.\ Phys.\ D {\bf 20} (2011) 413
doi:10.1142/S0218271811018925 [arXiv:1101.3864 [gr-qc]].



\bibitem{GBM:2017lvd}
  B.~P.~Abbott {\it et al.}
  ``Multi-messenger Observations of a Binary Neutron Star Merger,''
  Astrophys.\ J.\  {\bf 848} (2017) no.2,  L12
  doi:10.3847/2041-8213/aa91c9
  [arXiv:1710.05833 [astro-ph.HE]].




\bibitem{Ezquiaga:2017ekz}
  J.~M.~Ezquiaga and M.~Zumalacarregui,
  Phys.\ Rev.\ Lett.\  {\bf 119} (2017) no.25,  251304
  doi:10.1103/PhysRevLett.119.251304
  [arXiv:1710.05901 [astro-ph.CO]].


\bibitem{ArmendarizPicon:1999rj}
  C.~Armendariz-Picon, T.~Damour and V.~F.~Mukhanov,
  Phys.\ Lett.\ B {\bf 458} (1999) 209
  doi:10.1016/S0370-2693(99)00603-6
  [hep-th/9904075].


\bibitem{Chiba:1999ka}
T.~Chiba, T.~Okabe and M.~Yamaguchi,
Phys.\ Rev.\ D {\bf 62} (2000) 023511
doi:10.1103/PhysRevD.62.023511 [astro-ph/9912463].

\bibitem{ArmendarizPicon:2000dh}
C.~Armendariz-Picon, V.~F.~Mukhanov and P.~J.~Steinhardt,
Phys.\ Rev.\ Lett.\ {\bf 85} (2000) 4438
doi:10.1103/PhysRevLett.85.4438 [astro-ph/0004134].




\bibitem{Matsumoto:2010uv}
J.~Matsumoto and S.~Nojiri,
Phys.\ Lett.\ B {\bf 687} (2010) 236
doi:10.1016/j.physletb.2010.03.030 [arXiv:1001.0220 [hep-th]].



\bibitem{ArmendarizPicon:2000ah}
  C.~Armendariz-Picon, V.~F.~Mukhanov and P.~J.~Steinhardt,
  Phys.\ Rev.\ D {\bf 63} (2001) 103510
  doi:10.1103/PhysRevD.63.103510
  [astro-ph/0006373].


\bibitem{Chiba:2002mw}
  T.~Chiba,
  Phys.\ Rev.\ D {\bf 66} (2002) 063514
  doi:10.1103/PhysRevD.66.063514
  [astro-ph/0206298].



\bibitem{Malquarti:2003nn}
  M.~Malquarti, E.~J.~Copeland, A.~R.~Liddle and M.~Trodden,
  Phys.\ Rev.\ D {\bf 67} (2003) 123503
  doi:10.1103/PhysRevD.67.123503
  [astro-ph/0302279].



\bibitem{Malquarti:2003hn}
  M.~Malquarti, E.~J.~Copeland and A.~R.~Liddle,
  Phys.\ Rev.\ D {\bf 68} (2003) 023512
  doi:10.1103/PhysRevD.68.023512
  [astro-ph/0304277].



\bibitem{Chimento:2003zf}
  L.~P.~Chimento and A.~Feinstein,
  Mod.\ Phys.\ Lett.\ A {\bf 19} (2004) 761
  doi:10.1142/S0217732304013507
  [astro-ph/0305007].


\bibitem{Chimento:2003ta}
  L.~P.~Chimento,
  Phys.\ Rev.\ D {\bf 69} (2004) 123517
  doi:10.1103/PhysRevD.69.123517
  [astro-ph/0311613].


\bibitem{Scherrer:2004au}
  R.~J.~Scherrer,
  Phys.\ Rev.\ Lett.\  {\bf 93} (2004) 011301
  doi:10.1103/PhysRevLett.93.011301
  [astro-ph/0402316].


\bibitem{Aguirregabiria:2004te}
  J.~M.~Aguirregabiria, L.~P.~Chimento and R.~Lazkoz,
  Phys.\ Rev.\ D {\bf 70} (2004) 023509
  doi:10.1103/PhysRevD.70.023509
  [astro-ph/0403157].



\bibitem{ArmendarizPicon:2005nz}
  C.~Armendariz-Picon and E.~A.~Lim,
  JCAP {\bf 0508} (2005) 007
  doi:10.1088/1475-7516/2005/08/007
  [astro-ph/0505207].



\bibitem{Abramo:2005be}
  L.~R.~Abramo and N.~Pinto-Neto,
  Phys.\ Rev.\ D {\bf 73} (2006) 063522
  doi:10.1103/PhysRevD.73.063522
  [astro-ph/0511562].


\bibitem{Rendall:2005fv}
  A.~D.~Rendall,
  Class.\ Quant.\ Grav.\  {\bf 23} (2006) 1557
  doi:10.1088/0264-9381/23/5/008
  [gr-qc/0511158].


\bibitem{Bruneton:2006gf}
  J.~P.~Bruneton,
  Phys.\ Rev.\ D {\bf 75} (2007) 085013
  doi:10.1103/PhysRevD.75.085013
  [gr-qc/0607055].


\bibitem{dePutter:2007ny}
  R.~de Putter and E.~V.~Linder,
  Astropart.\ Phys.\  {\bf 28} (2007) 263
  doi:10.1016/j.astropartphys.2007.05.011
  [arXiv:0705.0400 [astro-ph]].


\bibitem{Babichev:2007dw}
  E.~Babichev, V.~Mukhanov and A.~Vikman,
  JHEP {\bf 0802} (2008) 101
  doi:10.1088/1126-6708/2008/02/101
  [arXiv:0708.0561 [hep-th]].


\bibitem{Deffayet:2011gz}
  C.~Deffayet, X.~Gao, D.~A.~Steer and G.~Zahariade,
  Phys.\ Rev.\ D {\bf 84} (2011) 064039
  doi:10.1103/PhysRevD.84.064039
  [arXiv:1103.3260 [hep-th]].


\bibitem{Kan:2018odq}
  N.~Kan, K.~Shiraishi and M.~Yashiki,
  arXiv:1811.11967 [gr-qc].







\bibitem{Unnikrishnan:2012zu}
  S.~Unnikrishnan, V.~Sahni and A.~Toporensky,
  JCAP {\bf 1208} (2012) 018
  doi:10.1088/1475-7516/2012/08/018
  [arXiv:1205.0786 [astro-ph.CO]].


\bibitem{Li:2012vta}
  S.~Li and A.~R.~Liddle,
  JCAP {\bf 1210} (2012) 011
  doi:10.1088/1475-7516/2012/10/011
  [arXiv:1204.6214 [astro-ph.CO]].







\bibitem{Inoue:2001zt}
S.~Inoue and J.~Yokoyama,
Phys.\ Lett.\ B {\bf 524} (2002) 15
doi:10.1016/S0370-2693(01)01369-7 [hep-ph/0104083].

\bibitem{Tsamis:2003px}
N.~C.~Tsamis and R.~P.~Woodard,
Phys.\ Rev.\ D {\bf 69} (2004) 084005
doi:10.1103/PhysRevD.69.084005 [astro-ph/0307463].

\bibitem{Kinney:2005vj}
W.~H.~Kinney,
Phys.\ Rev.\ D {\bf 72} (2005) 023515
doi:10.1103/PhysRevD.72.023515 [gr-qc/0503017].

\bibitem{Tzirakis:2007bf}
K.~Tzirakis and W.~H.~Kinney,
Phys.\ Rev.\ D {\bf 75} (2007) 123510
doi:10.1103/PhysRevD.75.123510 [astro-ph/0701432].

\bibitem{Namjoo:2012aa}
M.~H.~Namjoo, H.~Firouzjahi and M.~Sasaki,
Europhys.\ Lett.\  {\bf 101} (2013) 39001
doi:10.1209/0295-5075/101/39001 [arXiv:1210.3692 [astro-ph.CO]].

\bibitem{Martin:2012pe}
J.~Martin, H.~Motohashi and T.~Suyama,
Phys.\ Rev.\ D {\bf 87} (2013) no.2,  023514
doi:10.1103/PhysRevD.87.023514 [arXiv:1211.0083 [astro-ph.CO]].

\bibitem{Motohashi:2014ppa}
H.~Motohashi, A.~A.~Starobinsky and J.~Yokoyama,
JCAP {\bf 1509} (2015) no.09,  018
doi:10.1088/1475-7516/2015/09/018 [arXiv:1411.5021 [astro-ph.CO]].

\bibitem{Cai:2016ngx}
Y.~F.~Cai, J.~O.~Gong, D.~G.~Wang and Z.~Wang,
JCAP {\bf 1610} (2016) no.10,  017
doi:10.1088/1475-7516/2016/10/017 [arXiv:1607.07872
[astro-ph.CO]].

\bibitem{Motohashi:2017aob}
H.~Motohashi and A.~A.~Starobinsky,
arXiv:1702.05847 [astro-ph.CO].

\bibitem{Hirano:2016gmv}
S.~Hirano, T.~Kobayashi and S.~Yokoyama,
Phys.\ Rev.\ D {\bf 94} (2016) no.10,  103515
doi:10.1103/PhysRevD.94.103515 [arXiv:1604.00141 [astro-ph.CO]].

\bibitem{Anguelova:2015dgt}
L.~Anguelova,
Nucl.\ Phys.\ B {\bf 911} (2016) 480
doi:10.1016/j.nuclphysb.2016.08.020 [arXiv:1512.08556 [hep-th]].

\bibitem{Cook:2015hma}
J.~L.~Cook and L.~M.~Krauss,
JCAP {\bf 1603} (2016) no.03,  028
doi:10.1088/1475-7516/2016/03/028 [arXiv:1508.03647
[astro-ph.CO]].

\bibitem{Kumar:2015mfa}
K.~S.~Kumar, J.~Marto, P.~Vargas Moniz and S.~Das,
JCAP {\bf 1604} (2016) no.04,  005
doi:10.1088/1475-7516/2016/04/005 [arXiv:1506.05366 [gr-qc]].

\bibitem{Odintsov:2017yud}
S.~D.~Odintsov and V.~K.~Oikonomou,
arXiv:1703.02853 [gr-qc].

\bibitem{Odintsov:2017qpp}
S.~D.~Odintsov and V.~K.~Oikonomou,
arXiv:1704.02931 [gr-qc].

\bibitem{Lin:2015fqa}
J.~Lin, Q.~Gao and Y.~Gong,
Mon.\ Not.\ Roy.\ Astron.\ Soc.\  {\bf 459} (2016) no.4,  4029
doi:10.1093/mnras/stw915 [arXiv:1508.07145 [gr-qc]].

\bibitem{Gao:2017uja}
Q.~Gao and Y.~Gong,
arXiv:1703.02220 [gr-qc].



\bibitem{Nojiri:2017qvx}
  S.~Nojiri, S.~D.~Odintsov and V.~K.~Oikonomou,
  Class.\ Quant.\ Grav.\  {\bf 34} (2017) no.24,  245012
  doi:10.1088/1361-6382/aa92a4
  [arXiv:1704.05945 [gr-qc]].





\bibitem{Oikonomou:2017bjx}
  V.~K.~Oikonomou,
  Mod.\ Phys.\ Lett.\ A {\bf 32} (2017) no.33,  1750172
  doi:10.1142/S0217732317501723
  [arXiv:1706.00507 [gr-qc]].




\bibitem{Odintsov:2017hbk}
  S.~D.~Odintsov, V.~K.~Oikonomou and L.~Sebastiani,
  Nucl.\ Phys.\ B {\bf 923} (2017) 608
  doi:10.1016/j.nuclphysb.2017.08.018
  [arXiv:1708.08346 [gr-qc]].



\bibitem{Oikonomou:2017xik}
  V.~K.~Oikonomou,
  Int.\ J.\ Mod.\ Phys.\ D {\bf 27} (2017) no.02,  1850009
  doi:10.1142/S0218271818500098
  [arXiv:1709.02986 [gr-qc]].



\bibitem{Cicciarella:2017nls}
  F.~Cicciarella, J.~Mabillard and M.~Pieroni,
  JCAP {\bf 1801} (2018) no.01,  024
  doi:10.1088/1475-7516/2018/01/024
  [arXiv:1709.03527 [astro-ph.CO]].




\bibitem{Awad:2017ign}
  A.~Awad, W.~El Hanafy, G.~G.~L.~Nashed, S.~D.~Odintsov and V.~K.~Oikonomou,
  JCAP {\bf 1807} (2018) no.07,  026
  doi:10.1088/1475-7516/2018/07/026
  [arXiv:1710.00682 [gr-qc]].





\bibitem{Anguelova:2017djf}
  L.~Anguelova, P.~Suranyi and L.~C.~R.~Wijewardhana,
  JCAP {\bf 1802} (2018) no.02,  004
  doi:10.1088/1475-7516/2018/02/004
  [arXiv:1710.06989 [hep-th]].




\bibitem{Ito:2017bnn}
  A.~Ito and J.~Soda,
  Eur.\ Phys.\ J.\ C {\bf 78} (2018) no.1,  55
  doi:10.1140/epjc/s10052-018-5534-5
  [arXiv:1710.09701 [hep-th]].



\bibitem{Karam:2017rpw}
  A.~Karam, L.~Marzola, T.~Pappas, A.~Racioppi and K.~Tamvakis,
  JCAP {\bf 1805} (2018) no.05,  011
  doi:10.1088/1475-7516/2018/05/011
  [arXiv:1711.09861 [astro-ph.CO]].

\bibitem{Yi:2017mxs}
  Z.~Yi and Y.~Gong,
  JCAP {\bf 1803} (2018) no.03,  052
  doi:10.1088/1475-7516/2018/03/052
  [arXiv:1712.07478 [gr-qc]].





\bibitem{Mohammadi:2018oku}
  A.~Mohammadi, K.~Saaidi and T.~Golanbari,
  Phys.\ Rev.\ D {\bf 97} (2018) no.8,  083006
  doi:10.1103/PhysRevD.97.083006
  [arXiv:1801.03487 [hep-ph]].




\bibitem{Gao:2018tdb}
  Q.~Gao, Y.~Gong and Q.~Fei,
  JCAP {\bf 1805} (2018) no.05,  005
  doi:10.1088/1475-7516/2018/05/005
  [arXiv:1801.09208 [gr-qc]].





\bibitem{Mohammadi:2018wfk}
  A.~Mohammadi and K.~Saaidi,
  arXiv:1803.01715 [astro-ph.CO].




\bibitem{Morse:2018kda}
  M.~J.~P.~Morse and W.~H.~Kinney,
  Phys.\ Rev.\ D {\bf 97} (2018) no.12,  123519
  doi:10.1103/PhysRevD.97.123519
  [arXiv:1804.01927 [astro-ph.CO]].



\bibitem{Cruces:2018cvq}
  D.~Cruces, C.~Germani and T.~Prokopec,
  JCAP {\bf 1903} (2019) no.03,  048
  doi:10.1088/1475-7516/2019/03/048
  [arXiv:1807.09057 [gr-qc]].


\bibitem{GalvezGhersi:2018haa}
  J.~T.~Galvez Ghersi, A.~Zucca and A.~V.~Frolov,
  JCAP {\bf 1905} (2019) no.05,  030
  doi:10.1088/1475-7516/2019/05/030
  [arXiv:1808.01325 [astro-ph.CO]].



\bibitem{Boisseau:2018rgy}
  B.~Boisseau and H.~Giacomini,
  arXiv:1809.09169 [gr-qc].



\bibitem{Gao:2019sbz}
  Q.~Gao, Y.~Gong and Z.~Yi,
  arXiv:1901.04646 [gr-qc].


\bibitem{Lin:2019fcz}
  W.~C.~Lin, M.~J.~P.~Morse and W.~H.~Kinney,
  arXiv:1904.06289 [astro-ph.CO].









\bibitem{Liddle:2000cg}
  A.~R.~Liddle and D.~H.~Lyth,
  Cambridge, UK: Univ. Pr. (2000) 400 p





\bibitem{DeFelice:2011zh}
  A.~De Felice and S.~Tsujikawa,
  JCAP {\bf 1104} (2011) 029
  doi:10.1088/1475-7516/2011/04/029
  [arXiv:1103.1172 [astro-ph.CO]].




\bibitem{Noh:2001ia}
H.~Noh and J.~c.~Hwang,
Phys.\ Lett.\ B {\bf 515} (2001) 231
doi:10.1016/S0370-2693(01)00875-9 [astro-ph/0107069].

\bibitem{Hwang:2005hb}
J.~c.~Hwang and H.~Noh,
Phys.\ Rev.\ D {\bf 71} (2005) 063536
doi:10.1103/PhysRevD.71.063536 [gr-qc/0412126].


\bibitem{Hwang:2002fp}
J.~c.~Hwang and H.~Noh,
Phys.\ Rev.\ D {\bf 66} (2002) 084009
doi:10.1103/PhysRevD.66.084009 [hep-th/0206100].

\bibitem{Kaiser:2013sna}
D.~I.~Kaiser and E.~I.~Sfakianakis,
Phys.\ Rev.\ Lett.\ {\bf 112} (2014) no.1, 011302
doi:10.1103/PhysRevLett.112.011302 [arXiv:1304.0363
[astro-ph.CO]].




\bibitem{Lyth:2009zz}
  D.~H.~Lyth and A.~R.~Liddle,
  Cambridge, UK: Cambridge Univ. Pr. (2009) 497 p


\bibitem{Stewart:1993bc}
  E.~D.~Stewart and D.~H.~Lyth,
  Phys.\ Lett.\ B {\bf 302}, 171 (1993)
  doi:10.1016/0370-2693(93)90379-V
  [gr-qc/9302019].


\bibitem{Habib:2002yi}
  S.~Habib, K.~Heitmann, G.~Jungman and C.~Molina-Paris,
  Phys.\ Rev.\ Lett.\  {\bf 89} (2002) 281301
  doi:10.1103/PhysRevLett.89.281301
  [astro-ph/0208443].



\bibitem{Oikonomou:2017isf}
  V.~K.~Oikonomou,
  Phys.\ Rev.\ D {\bf 95} (2017) no.8,  084023
  doi:10.1103/PhysRevD.95.084023
  [arXiv:1703.10515 [gr-qc]].


\bibitem{Keskin:2018gev}
  A.~I.~Keskin,
  Eur.\ Phys.\ J.\ C {\bf 78} (2018) no.9,  705.
  doi:10.1140/epjc/s10052-018-6199-9




\bibitem{Akrami:2019izv}
  Y.~Akrami {\it et al.} [Planck Collaboration],
  arXiv:1905.05697 [astro-ph.CO].










\end{thebibliography}
\end{document}